\documentclass[twocolumn,showpacs,preprintnumbers,amsmath,amssymb]{revtex4}

\usepackage{epsfig,psfrag}
\usepackage{dcolumn}
\usepackage{bm}

\begin{document}


\title{ Destruction of interference by
many-body interactions in cold atomic Bose gases  }

\author{S.~Chen and R.~Egger}

\affiliation{Institut f\"ur Theoretische Physik, 
Heinrich-Heine-Universit\"at,
 D-40225 D\"usseldorf, Germany}

\date{\today}

\begin{abstract}
We study the effects of many-body interactions on the interference
in a Mach-Zehnder interferometer for matter
waves of ultracold Bose atoms.  After switching off
an axial trapping potential, the thermal
initial wavepacket expands, and subsequently interference fringes
may be observed in a circular 1D trap.  These are computed for
axial harmonic or $\delta$-function traps, and for 
interaction strengths from the Thomas-Fermi regime to
the Tonks-Girardeau limit.  It is shown that many-body correlations
in a realistic setup destroy interference to a large degree.
Analytical expressions allowing to infer
the observability of phase coherence and interference are provided.
\end{abstract}

\maketitle

\section{Introduction}

Recent experimental advances in the control of ultracold atomic gases 
have opened new exciting possibilities to systematically study the effects
of many-body interactions in  low-dimensional 
strongly correlated mesoscopic systems \cite{Kasevich,Feshbach,Rolston}. 
Two main experimental routes are
presently being pursued, namely magnetic or magneto-optical trap
technology \cite{Rolston,Dumke,Schreck,Richard}, and the formation 
of atomic waveguides on 
microchips \cite{Hansel,Ott,Pritchard,Bongs}.   
A particularly interesting limit arises in the true 1D case,
which may, for instance,
 be realized by confining $N$ atoms in a highly anisotropic cigar-shaped
trap with large transverse trapping frequency $\omega_\perp$
\cite{Gorlitz,Dettmer}.   
In this paper we study interacting
1D Bose gases, but similar ideas and conclusions apply to cold fermionic
gases as well.  In the 1D case, in an infinitely long system,
quantum fluctuations prevent the existence of a Bose-Einstein
condensate (BEC) even at $T=0$.  For a finite system and very
weak interactions, however, BEC could happen 
at extremely low temperatures not reachable at present 
\cite{Monien,Petrov1D,Lieb2,Graham1,Graham2}.
The physics of a trapped 1D atomic Bose gas
is then characterized by a crossover from the high-density  
weakly correlated Thomas-Fermi (TF) 
regime to  a strongly  correlated Tonks-Girardeau (TG) regime 
\cite{Tonks,Gir1} at low density, 
see Refs.~\cite{Petrov1D,Dunjko,mora,stoof}.  
The relevant dimensionless  interaction
parameter governing this crossover in a uniform
system is 
\begin{equation} \label{gamma} 
\gamma= mg/\hbar^2 \rho_0,
\end{equation}
 where $g$ is an effective 1D
interaction constant \cite{Olshanii},  see Eq.~(\ref{gdef}) below,
 and $\rho_0$ is the density.
While for the uniform system, a Bethe ansatz solution due to Lieb and
Liniger \cite{Lieb} is available, for the case of a
(possibly time-dependent) axial trapping potential $V(x,t)$,
no exact statements are known, and one generally 
resorts to approximate or numerical methods \cite{GW,Ohberg,Pedri}.  
We mention in passing that this
system also has very interesting
 single-particle excitations with exotic properties
such as fractional (anyon) statistics.  
While such systems could also be realized in more conventional
condensed matter systems,  the high quality and tunability
of atomic Bose gases renders them very attractive
for fundamental studies. 

\begin{figure}
\centerline{\epsfysize=8cm \epsffile{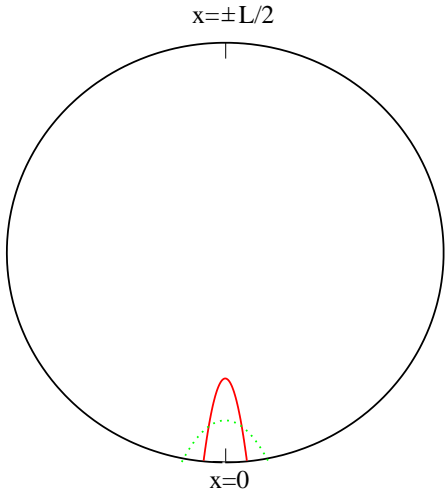}}
\caption{\label{fig1} Schematic geometry of a Mach-Zehnder
interferometer realized in a circular trap 
 with circumference $L$.  By applying an axial trap potential
at $t<0$ around $x\approx 0$, a thermal non-equilibrium initial state
is prepared, whose time evolution and interference
at $x\approx \pm L/2$ are the subject of this paper.
}
\end{figure}

In this paper, we address the influence of interactions on the 
phase coherence and the interference properties of a 1D Bose gas.
To be specific, we consider the 
circular geometry indicated in Fig.~\ref{fig1}.  At time $t=0$,
the axial trapping potential $V(x,t)$ localized around $x\approx 0$ is 
switched off, and  after some characteristic expansion time, the
expanding right- and left-moving wave packets will then meet at the other side
of the ring, $x\approx \pm L/2$.
If phase coherence of the wave packets is not lost during the
propagation, a Mach-Zehnder-type
 interference signal should be observable at this side.
While a modified Gross-Pitaevskii  (GP)
 approach, obtained from combining the Lieb-Liniger solution with 
a local density approximation (LDA) 
valid for large enough systems \cite{Dunjko},
 is able to yield
accurate ground-state density profiles $\rho_0(x,t)$ 
of a 1D Bose gas after switching off the axial trap even 
in the Tonks limit \cite{Ohberg},
it typically overestimates interference signals
by orders of magnitude \cite{GW}.
For a correct description of
interference, it is thus necessary to take into account quantum-mechanical
phase fluctuations, which have so far
only been addressed in the static case \cite{Gangardt,Petrov3D}
or in the Tonks limit under rather special initial conditions
\cite{Das1,Das2}. 
Below we show that under realistic modelling of the initial preparation
and the subsequent dynamics,
phase coherence and interference are strongly reduced by 
many-body interactions in 1D Bose gases. Moreover,
we provide quantitative expressions
to assess this effect in experiments. 
Proper insight into this question is of practical importance for
the development of future atom interferometers and for the
understanding of current experiments \cite{Pfau,Hellweg,Bongs}.
The remainder of the paper is organized as follows.
In Sec.~\ref{sec2} we study magneto-optical or magnetic
traps with 
\begin{equation}\label{trappot}
V(x,t)= \frac12 m\omega_x^2 x^2 \Theta(-t),
\end{equation}
with angular frequencies $\omega_x\ll \omega_\perp$; here
 $\Theta$ is the Heaviside function.
In Sec.~\ref{sec2}, the validity of LDA is presumed.
In Sec.~\ref{sec3},  we briefly discuss a related setup employing
a $\delta$-trap which
may be more realistic for microchip-trapped gases yet
allows for an exact solution of the time-dependent problem.
Finally, we conclude in Sec.~\ref{sec4}.

\section{Interference signal: Harmonic initial trap}
\label{sec2}

Let us consider a circular trap 
at ultralow temperature, $k_B T\ll
\hbar \omega_\perp$, such that motion in the transverse direction is frozen out,
see Fig.~\ref{fig1}.
The effective 1D interaction strength is \cite{Olshanii}
\begin{equation}\label{gdef}
g= -\frac{2\hbar^2}{m a_{\rm 1D}} , \quad
a_{\rm 1D}= - \frac{ d_\perp^2}{2a_s} \left[1-{\cal C}(a_s/d_\perp)\right],
\end{equation}
where $a_s>0$ is the 3D $s$-wave scattering length
describing the repulsive atom-atom interactions,
$d_{\perp} = (2\hbar/m \omega_{\perp})^{1/2}$, and
 ${\cal C}\approx 1.4603$.  
The interaction strength can be tuned by Feshbach resonances \cite{Inouye} or
by using optical lattices \cite{Feshbach}.

\subsection{Fluctuation modes}

 The resulting
Hamiltonian of the 1D Bose gas at fixed particle number $N$,
including an arbitrary time-dependent axial trap potential $V(x,t)$, is
\begin{eqnarray} \nonumber
H(t) & = & \int_{-L/2}^{L/2}
 d x \ {\hat{\psi}}^{\dagger}(x,t) \left[ -\frac{\hbar^2}{2m}  \partial_x^2 
+V(x,t) \right]\hat{\psi}(x,t) \\ & +& \frac{g}{2} \int d x 
: \left[ {\hat{\psi}}^{\dagger}(x,t) \hat{\psi}(x,t) \right]^2 :  ,
 \label{1dH}
\end{eqnarray}
where the colons denote normal-ordering.
For $V(x,t)=0$, this model was solved exactly by Lieb and Liniger \cite{Lieb}. 
To make progress in the presence of the
external potential, the field operator $\hat{\psi}$ is expressed
in terms of the density $\rho(x)$ and the phase $\phi(x)$, 
\begin{equation} \label{density-phase}
\hat{\psi}(x) = \sqrt{\rho(x)} e^{i\phi(x)} .
\end{equation}
We employ canonically conjugate density [phase]
fluctuation operators $\Pi(x,t)=\rho(x,t)-\rho_0(x,t)$
[$\Phi(x,t) =\phi(x,t)-\phi_0(x,t)$] describing 
quantum fluctuations around the solution  
 $ \psi_0(x,t) = \sqrt{\rho_0(x,t)} \exp[i\phi_0(x,t)]$  
of the time-dependent modified GP equation \cite{Dunjko,Ohberg},
\begin{equation} \label{TGPE}
i \hbar \partial_t \psi_0(x,t)= 
\left[ - \frac{\hbar^2}{2 m} \partial_x^2 + V(x,t)
+ \tilde{F}(\rho_0(x,t)) \right] \psi_0(x,t),
\end{equation} 
where $\tilde{F}(\rho_0)$ is the (exactly known)
chemical potential for the Lieb-Liniger problem
\cite{Dunjko}, with limiting values 
\begin{equation}\label{frho}
\tilde{F}(\rho_0)= \left \{ \begin{array}{cc} 
g\rho_0, & \gamma \ll 1\\
\pi^2\hbar^2 \rho_0^2/2m, & \gamma \gg 1. \end{array} \right.
\end{equation}
For arbitrary time-dependent potential $V(x,t)$,  
no closed solution of Eq.~(\ref{TGPE}) for arbitrary $\gamma$ is available. 
{}From numerical work, however,
the expansion of a 1D Bose gas after switching off an harmonic axial trap
potential has recently been shown to violate self-similarity \cite{Ohberg} 
in the intermediate regime between the TF and the TG limit, $\gamma\approx 1$,
which can be traced back to the variable nonlinearity of $\tilde{F}(\rho_0)$ for
different $\gamma$.

Substituting Eq.~(\ref{density-phase}) into
Eq.~(\ref{1dH}) and expanding up to quadratic order
in $\Phi$ and $\Pi$,  we obtain  the effective low-energy
 Hamiltonian describing quantum fluctuations around the
solution to Eq.~(\ref{TGPE}).  
Neglecting terms containing $\Pi\partial_x^2 \Pi$ or 
$(\rho_0^{-1}\partial^2_x \rho_0)\Pi^2$, we obtain
the time-dependent Hamiltonian 
\begin{eqnarray} \label{HLLt}
H(t) & = & \int dx \Bigl [\frac{\hbar^2 \rho_0}{2m}
(\partial_x\Phi)^2 \\ \nonumber &+& \frac{\hbar^2}{m} (\partial_x\phi_0)
\Pi\ \partial_x \Phi + \frac{1}{2}
\frac{\partial \tilde{F}(\rho_0)}{\partial\rho_0} \Pi^2  \Bigr]  .
\end{eqnarray}
While terms involving $\Pi \partial_x^2\Pi$ are irrelevant in
the long-wavelength low-energy regime, the neglect of terms $\propto
(\rho_0^{-1}\partial_x^2 \rho_0) \Pi^2$ requires that the interaction
strength $g$ is not too weak \cite{Wu}.  In practice, this
approximation is valid to high accuracy even in the TF regime,
and therefore in all cases of interest here.
Interference fringes are correctly captured in such
a long-wavelength theory despite their rapidly oscillating character,
 since the oscillations arise due to a mixing
of left- and right-moving wave-packets. These wave-packets are both
correctly described within the present theory.

Equation (\ref{HLLt}) resembles
 the Hamiltonian of a Luttinger liquid \cite{Monien},
generalized to a time-dependent and spatially non-uniform system,
where a non-standard coupling between $\Pi$ and $\partial_x \Phi$ 
is induced by the time dependence of the external potential.
The Hamiltonian (\ref{HLLt}) is quadratic in the fluctuation operators and 
can therefore be diagonalized by a Bogoliubov-de Gennes transformation.
However, explicit diagonalization is rather difficult for 
arbitrary potential $V(x,t)$, and here we shall limit ourselves
to the form given in Eq.~(\ref{trappot}).
Although the full Hamiltonian (\ref{1dH}) is time-independent for $t>0$,
the Hamiltonian (\ref{HLLt}) governing quantum fluctuations becomes
explicitly time-dependent due to the time dependence of
$\rho_0(x,t)$ and $\phi_0(x,t)$.
Furthermore, to make progress, we restrict ourselves to the 
self-similar limits $\gamma\ll 1$ and $\gamma\gg 1$. Then the 
density $\rho_0(x,t>0)$ is self-similar with some 
scale factor $b(t)$ \cite{Ohberg}, 
\begin{equation}\label{selfsimilar}
\rho_0(x,t) = \frac{n(x/b(t))}{b(t)},
\end{equation}
where $b(0)=1$ and $n(x)=\rho_0(x,0)$ is the initial density profile.
Solving Eq.~(\ref{TGPE}) with Eq.~(\ref{selfsimilar}) results in 
the phase 
\begin{equation}\label{phi0}
\phi_0(x,t) = \frac{mx^2}{2\hbar} \frac{\dot{b}(t)}{b(t)} .
\end{equation}

Under self-similarity, it is then possible to diagonalize the
quadratic Hamiltonian (\ref{HLLt}) by 
solving the equation of motion generated by it.  
Using the long derivative
\begin{equation}\label{lder}
\hat{ D } = \partial_t + x (\dot{b}/b) \partial_x ,
\end{equation}
some algebra shows that the equation of motion for $\Phi$ is
\begin{equation}\label{f_jLL}
\left (\hat{D}+ \alpha \frac{\dot{b}}{b}\right) \hat{D}
 \Phi(x,t) =\frac{1}{m}  \frac{\partial \tilde{F}(\rho_0)}{\rho_0}
\partial_x [ \rho_0 \partial_x\Phi],
\end{equation}
where $\alpha=1$ in the TF regime and $\alpha=2$ in the TG regime.
To solve Eq.~(\ref{f_jLL}), we use the ansatz 
\begin{equation}\label{phians}
\Phi(x,t) = \sum_j  C_j(t) f_j(x/b(t)) \hat{B}_j(t) + {\rm H.c.},
\end{equation}
where $\hat{B}_{j}^{\dagger },\hat{B}_{j}$ are standard Bose operators with
time dependence governed by time-dependent eigenfrequency $\Omega_j(t)$,
\begin{equation}
\hat{B}_j(t)= \exp\left [-i\int_0^t dt' \Omega_j(t') \right] \hat{B}_j(0).
\end{equation}
Furthermore, $f_j$ are suitably normalized eigenfunctions, 
and the prefactor $C_j$ is needed to bring $H(t)$ into the 
canonical form
\begin{equation}\label{hom}
H (t) = \sum_{j}\hbar \Omega_j (t) 
\hat{B}_{j}^{\dagger}(t) \hat{B}_{j}(t),
\end{equation}
where the summation extends over all eigenmodes.
One finds a similar expression to Eq.~(\ref{phians})
 for the conjugate field $\Pi$
by using the commutation relation with $\Phi$.  
Given the scale-factor $b(t)$ and the initial density $n(x)$ 
entering Eq.~(\ref{selfsimilar}),
solving the eigenproblem (\ref{f_jLL}) thus leads to a complete
description of the time-dependent quantum phase fluctuations.
This program is carried out in this
section separately for the TF limit, see Sec.~\ref{sec22},
 and the TG regime, see Sec.~\ref{sec23}.

To investigate coherence properties during the self-similar 
expansion process, we calculate the (equal-time) density matrix 
\begin{equation}\label{1pdm}
W(x,x',t)=\langle \hat{\psi}^{\dagger}(x,t) \hat{\psi}(x',t) \rangle.
\end{equation}
Following Refs.~\cite{Ho,Petrov1D}, 
here it is justified to neglect density fluctations
against phase fluctuations, resulting in 
\begin{eqnarray} \label{W}
W(x, x', t) &\simeq& W_0(x,x',t) e^{-F(x,x',t)}, \\ \nonumber
W_0(x,x',t) &=& \sqrt{\rho_0(x,t)\rho_0(x',t)}
e^{-i[\phi_0(x,t)-\phi_0 (x',t)]}, \\ \nonumber
F(x,x',t) & =&  \langle [\Phi(x,t)-\Phi(x',t)]^{2} \rangle/2.
\end{eqnarray}
The task is then to compute the fluctuation correlator $F$.  Substituting 
Eq.~(\ref{phians}) into $F(x,x',t)$, it follows that
\begin{eqnarray}\nonumber
F(x,x^{\prime },t)  
& = & \sum _{j=0}^{\infty}
 \frac12 |C_j(t)|^2  (1 + 2 n_j) 
  [ f_{j}(x,t) - f_j(x',t) ]^2, \\
& = & \sum _{j=0}^{\infty}
 |C_j(t)|^2  \coth[\hbar \Omega_j(t)/2 k_B T] \nonumber \\
 \label{F}
 &\times&
 \frac12 [ f_{j}(x,t) - f_j(x',t) ]^2,
\end{eqnarray}
with the Bose-Einstein distribution
 $n_j = \langle \hat{B}_{j}^{\dagger} \hat{B}_{j} \rangle
= [\exp(\hbar \Omega_j(t)/ k_B T) -1]^{-1}$.

To analyze interference fringes, it is convenient to
switch to $y=x- L/2$ ($y=x+L/2$) for $L/2>x>0$ $(-L/2<x<0)$, such that the
center of the overlapping clouds is at $y=0$,
and the detection signal is given by
\begin{eqnarray*} 
I(y,t) &=& \langle\rho(y+L/2,t) \rangle + \langle\rho(-y+L/2,t)\rangle \\ 
& +& 2 {\rm Re}\ W(y+L/2,-y+L/2,t) .
\end{eqnarray*}
The last term corresponds to the (first-order) interference  signal
$I_{\rm int}(y)$ of interest here.  Using Eq.~(\ref{W}), 
it reads
\begin{eqnarray} \label{signal}
  I_{\rm int}(y) &=& 2 \cos[2 Q y] e^{-F(y+L/2,- y+L/2,t)} \\
\nonumber
&\times& \sqrt{\rho_0 (y+L/2,t)\rho_0(- y+L/2,t)} ,
\end{eqnarray}
with time-dependent  modulation wavevector $Q= [\phi_0(y+L/2,t)-\phi_0(-y
+L/2,t)]/2y$, which for $|y|\ll L$ takes the form
\begin{equation}\label{Q}
Q(t) =  \partial_x\phi_0(L/2,t) = \frac{m L}{2\hbar} \frac{\dot{b}}{b}.
\end{equation} 
It is obvious from Eq.~(\ref{signal}) that
phase fluctuations entering $F(x,x',t)$ suppress interference.

\subsection{Thomas-Fermi regime}\label{sec22}

In this subsection, we assume that the initial trapped gas (at $t<0$)
is within the TF regime, and also stays within this 
regime throughout all relevant timescales. 
Then the initial density is given by \cite{Kagan}
\begin{equation} \label{rho_TF}
n(x) = \frac{m}{2g} \omega_x^2 (R^2 -x^2) \Theta(R^2-x^2),
\end{equation}
where the TF radius $R$ follows from
$\int n(x) dx = N$, 
\begin{equation}\label{rTF}
R = (3 N l_x^4 / |a_{\rm 1D}|)^{1/3}, \quad
l_x =\sqrt{\hbar/m \omega_x}.
\end{equation}
Provided the expanding gas indeed stays in the TF regime,
the expansion is well described by the
self-similar profile (\ref{selfsimilar}), 
where $b(t)$ is the solution to \cite{Kagan,Castin}
\begin{equation}\label{tfb}
\ddot{b} =\omega_x^2/b^2 , \quad {\rm with} \ 
b(0)=1, \ \dot{b}(0) =0.
\end{equation}
The phase $\phi_0(x,t)$ is then given by Eq.~(\ref{phi0}).
While the eigenproblem for the quantum
fluctuations has previously been solved in the static TF case
 \cite{Stringari,Wu,Ho,Petrov1D,Luxat}, the
coupling of $\partial_x \Phi$ and $\Pi$ in the time-dependent case, see
Eq.~(\ref{HLLt}), complicates the analysis considerably.
Remarkably, it is nevertheless possible to solve the full dynamical problem
analytically.

The equation of motion (\ref{f_jLL}) is solved under the 
ansatz (\ref{phians}) with 
\begin{equation}
C_j(t) = [2\hbar R \Omega_j(t) b(t)/g ]^{-1/2}.
\end{equation}
To see this, we note that the long derivative (\ref{lder}) 
acts on the part belonging to the annihilation operator
$\hat{B}_j$ in Eq.~(\ref{phians}) as 
\[
\hat {D} \Phi_j =\left (-i\Omega_j -\frac{\dot{b}}{2b}
-\frac{\dot{\Omega_j}}{2\Omega_j} \right) \Phi_j.
\]
After some algebra, using $y=x/R b(t)$, Eq.~(\ref{f_jLL}) then takes the form
\begin{equation}\label{legendre}
 \frac{G_j}{\omega_x^2}
 f_j(y) +  \frac{d}{dy} \left [(1-y^2) \frac{d}{dy} f_j(y) \right]=0,
\end{equation}
where $G_j(t)$ is given by
\begin{equation}\label{Gj}
G_j = b^3 \left(\Omega_j^2 - \frac{3}{4} (\dot{\Omega}_j/\Omega_j)^2
+ \ddot{\Omega}_j/2\Omega_j - (\dot{b}/2b)^2 
+ \ddot{b}/2b\right) .
\end{equation}
Equation (\ref{legendre}) has a standard solution in terms
of Legendre polynomials $P_j$, with normalized eigenfunctions
\begin{equation}
f_j(y)=  \sqrt{j+1/2} \ P_j(y), \quad \int_{-1}^{1} dy F^2_j(y)=1,
\end{equation}
iff $G_j(t)$ is time-independent and given by 
\begin{equation}\label{deq}
G_j[\Omega_j(t)] = j(j+1) \omega_x^2 ,
\end{equation}
where $j$ is a non-negative integer. 
Eigenfrequencies $\Omega_j(t)$  thus follow as solutions
to the differential equation induced by Eq.~(\ref{deq}).
On not too long time scales,
Eq.~(\ref{deq}) with (\ref{Gj}) has the solution 
\begin{equation} \label{omegajt}
\frac{\Omega_j(t)}{\omega_x} = \sqrt{\frac{2j(j+1)+1}{4 b^3(t)} } , 
\end{equation}
which holds as long as $\dot{b}(t)/2b(t) \ll \Omega_j(t)$. 
For the results shown below in Figs.~\ref{fig2} and \ref{fig3}, this
condition is accurately fulfilled. 
 On much longer time scales, one should instead
(numerically) solve Eq.~(\ref{deq}).
The fluctuation correlations (\ref{W}) and the interference 
signal (\ref{signal}) then follow immediately using these results.
It is obvious that $C_j(t) \propto g^{1/2}$,  and therefore the 
interaction parameter
$g$ enters the fluctuation correlator $F$ and 
suppresses the interference in an exponential manner.
The validity of the self-similar TF solution 
and the subsequent treatment of the quantum fluctuation modes 
is guaranteed by the validity of $\tilde{F}(\rho_0)=g\rho_0$, which in turn is 
accurate to $1\%$  for $\gamma<0.05$  \cite{Lieb}. 

To give concrete examples, consider a system of $N=10^3 \ $  ${}^{23}$Na
atoms on a ring of circumference $L=16 R$, where $R$ is the
TF radius.   
The axial trap potential is switched off at $t=0$, and we study the
resulting interference signal.
A qualitative estimate for the effective interaction strength 
at time $t>0$ can be
given in terms of $\gamma(t)=2/[|a_{1D}|\rho_m(t)]$
\cite{Dunjko,Ohberg}, where $\rho_m(t)=\rho_0(x=0,t)$.
Let us first consider a system with trap frequencies $\omega_x=0.5$~kHz
and $\omega_\perp=50$~kHz, where $\gamma(0) \approx 0.0006$ 
indicates that the system is deeply in the TF regime
and phase fluctuations are very small.
The expanding wave packets begin to meet at time $t= 13$~ms. 
We calculated the phase fluctuations from Eq.~(\ref{F}) 
with a UV cutoff $\hbar \Omega_j < \mu$  given by 
the $t=0$ TF chemical potential, $\mu=m\omega_x^2 R^2/2$,
see Eq.~(\ref{rTF}).
In Fig.~\ref{fig2}, the resulting 
interference fringes at time $t=$~16 ms  
corresponding to $b(t) = 10$ are shown.
While there is a detectable interference signal at $T=4$~nK, interference
is completely washed out by thermal fluctuations at 10 nK.
As second example, consider the same system but with higher trap
frequencies,
$\omega_x = 1$~kHz and $\omega_{\perp}=100$~kHz,
see Fig.~\ref{fig3}.
 Here we have stronger interactions, $\gamma(0) \approx 0.001$, but
the system still stays in  the TF regime on the timescales of interest.
Clearly, pronounced interference patterns can be observed again.

\begin{figure}
\centerline{\epsfysize=7cm \epsffile{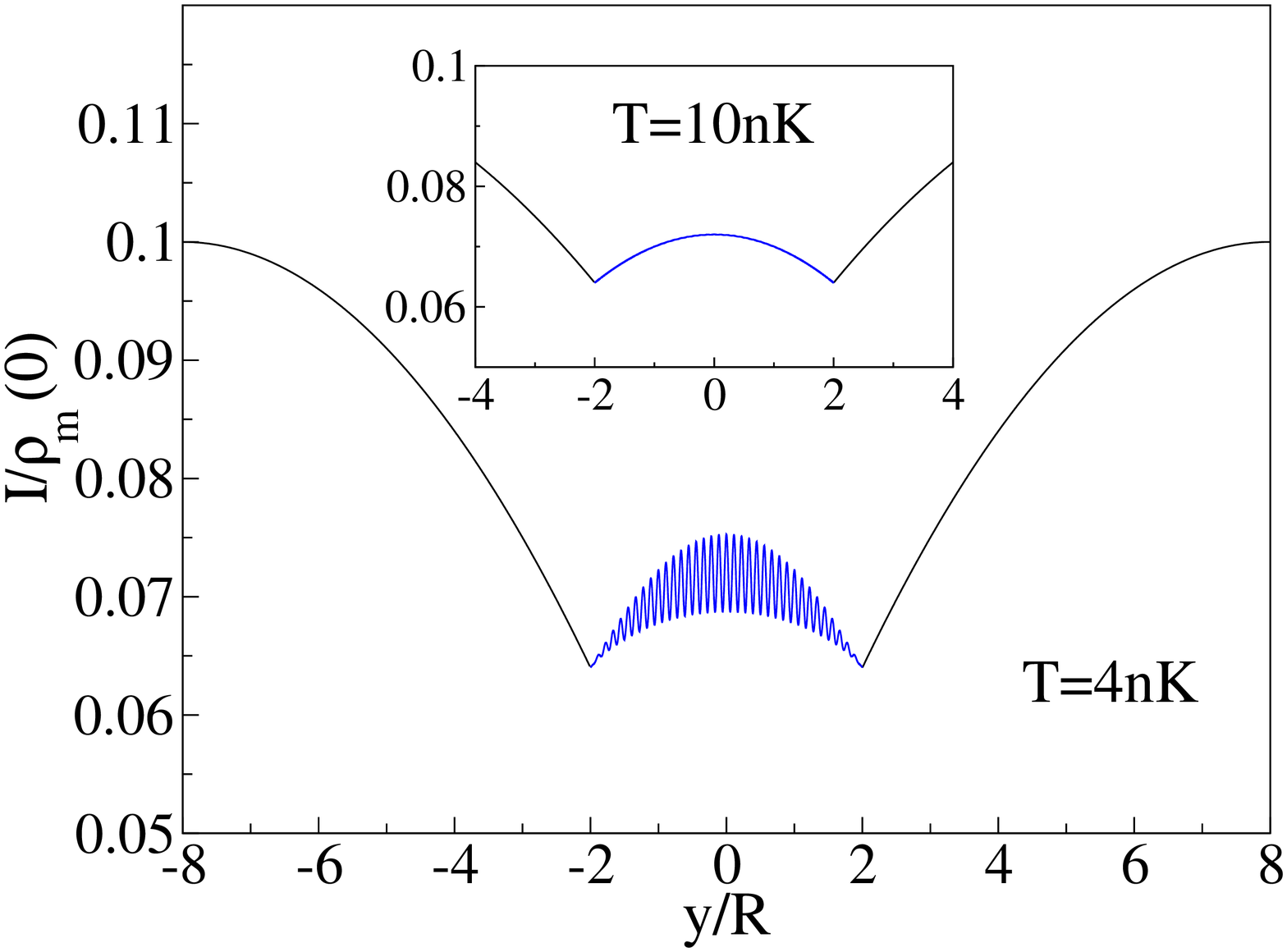}}
\caption{ \label{fig2}
 The scaled density profile in units of 
$\rho_m(0)$ for $N=10^3$ expanding Na atoms at time 
$t=16$~ms and temperature $T=4$ nK. 
Trap frequencies are $\omega_x = 0.5$~kHz and 
$\omega_{\perp}=50$~kHz. The inset shows the
corresponding density profile at $T=10$~nK.}
\end{figure}

\begin{figure}
\centerline{\epsfysize=7cm \epsffile{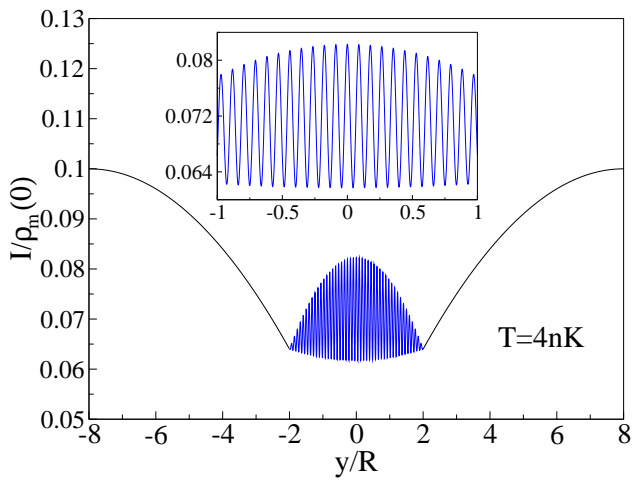}}
\caption{ \label{fig3}  Same as Fig.~\ref{fig2}, but with 
$\omega_x=1$~kHz and $\omega_\perp=100$~kHz.  The expanding clouds
meet at time $t=6.5$~ms, the snapshot is shown for $t=8$~ms.
The inset shows details of the interference fringes.}
\end{figure}

\subsection{Tonks-Girardeau regime}\label{sec23}

When the system leaves the TF regime during the expansion process,
self-similarity is violated \cite{Ohberg}, and
the solution of Eq.~(\ref{f_jLL}) is much more complicated.  
However, analytical progress
is possible in the opposite TG limit. Assuming that one starts
with an initial trap corresponding to the TG regime, during the
subsequent expansion one will also stay in this regime,
and the density becomes self-similar again \cite{Dunjko}, 
see Eq.~(\ref{selfsimilar}) with  $b(t)=\sqrt{1+\omega_x^2 t^2}$ and
\begin{equation}\label{tonks}
n(x)= n_{\rm Tonks}^0 \sqrt{1-x^2/R_0^2},
\end{equation} 
where $n_{\rm Tonks}^0=2N/\pi R_0$ and
\begin{equation}
R_0= \sqrt{2N\hbar/m\omega_x}
\end{equation}
denotes the Tonks radius. Furthermore, Eq.~(\ref{phi0}) specifies
$\phi_0(x,t)$ again.

Now Eq.~(\ref{f_jLL}) can be solved by the ansatz (\ref{phians}) with
\begin{equation}\label{ans2}
C_j(t) = [2\hbar m R_0 \Omega_j(t) b^2(t)/ \pi^2\hbar^2
 n_{\rm Tonks}^0]^{-1/2}.
\end{equation}
Proceeding in the same spirit as above,  using $y=x/R_0 b(t)$,
Eq.~(\ref{f_jLL}) leads to  
\begin{equation}\label{fff}
\frac{ \tilde{G}_j}{\omega_x^2}
 f_j(y) +  (1-y^2) \frac{d^2}{dy^2} f_j(y) - y \frac{d}{dy} f_j(y) =
0,
\end{equation}
where 
\begin{equation}\label{tGj}
\tilde{G}_j  =  b^4 \left(\Omega_j^2 - \frac{3}{4}
(\dot{\Omega}_j/\Omega_j)^2 + \ddot{\Omega}_j/2\Omega_j 
+ \ddot{b}/b \right) .
\end{equation}
With $j$ denoting positive integers, Eq.~(\ref{fff}) has
a solution provided that
\begin{equation}\label{deq2}
\tilde{G}_j = j^2 \omega_x^2,
\end{equation}
which gives a differential equation determining $\Omega_j(t)$.
The solution to Eq.~(\ref{fff}) is then
\begin{equation}\label{tonksnew}
f_j(y) =\sqrt{ \frac{2j\Gamma(j+1)\Gamma(j)}{\Gamma^2(j+1/2)}} 
P_j^{(-1/2, -1/2)} (y) ,
\end{equation}
where $P_j^{(-1/2,-1/2)}$ are Jacobi polynomials, and the
normalization condition is 
\[
\int_{-1}^{1} dy (1-y^2)^{-1/2} f_j^2(y)=1.
\]
Amazingly, Eq.~(\ref{deq2}) is solved by the simple form
\begin{equation}\label{om2}
\Omega_j(t)/\omega_x =  j/b^2(t),
\end{equation}
which holds on all timescales.
It is then straightforward 
to obtain the fluctuation correlations from Eq.~(\ref{F}), 
and the interference signal from Eq.~(\ref{signal}). 

\begin{figure}
\centerline{\epsfysize=7cm \epsffile{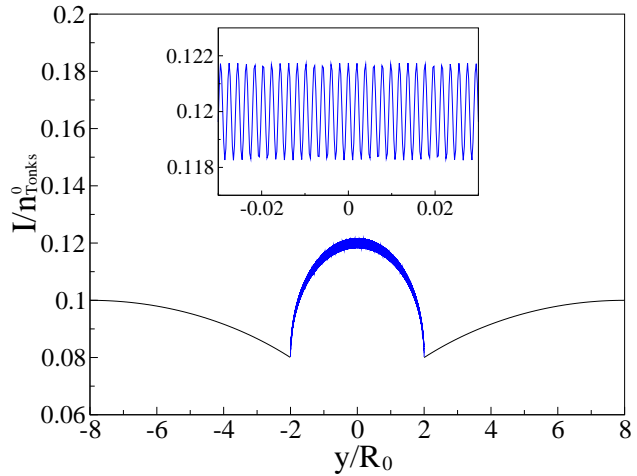}}
\caption{ \label{fig4}
The scaled density profile in units of $n_{\rm Tonks}^0$
for $N=10^3$ expanding ${}^{87}$Rb atoms at time $t=1$~s (see text).
The inset shows details of the interference fringes.
}
\end{figure}

Notably, even at zero temperature, interference is practically
completely suppressed by many-body interactions.
In Fig.~\ref{fig4}, we show the results of the above
calculation for $N=10^3 \ $  ${}^{87}$Rb
 atoms on a ring of circumference $L=16 R_0$, where
we assume a scattering length $a_s= 42.32$~nm,
e.g.~resulting from tuning via a Feshbach resonance. 
Trap frequencies are $\omega_x = 10$~Hz and 
$\omega_{\perp}=100$~kHz, which puts the system well into
the Tonks regime, $\gamma(0)\approx 14.3$.
At $t=0.8$~s the expanding wavepackets begin to meet, and
the scaled density profile 
at time $t=1$~s and zero temperature is shown in Fig.~\ref{fig4}. 
Interference amplitudes have decreased by orders of magnitude
when compared to the TF regime, but still tiny signatures
are observable.
At finite temperature, one then finds virtually no
sign of interference anymore, in accordance with 
expectations based on previous work \cite{Gangardt}.
As a result,  phase coherence is drastically suppressed in 
the Tonks limit.  

Finally, in between the TF and TG limits, one must resort to 
numerical techniques.  After numerical solution of Eq.~(\ref{TGPE}),
numerical data for $\rho_0(x,t)$ and $\phi_0(x,t)$
 could be used to solve the 
eigenproblem corresponding to Eq.~(\ref{f_jLL}),
and subsequently the interference
problem.  It is presently unclear whether this program can be
carried out in practice.
We shall pursue a different route in the next section, where analytical
results are given for a related setup employing a 
$\delta$ potential for the initial axial trap. This allows to access
the full crossover between the two limiting cases
 without approximations.
These results qualitatively confirm our 
above statements obtained in the
TF and TG limits.

\section{Alternative approach}
\label{sec3}

So far we have studied interference properties of atomic Bose gases
in 1D circular traps
after switching off an axial harmonic trap potential.  Our
treatment had to rely on 
the local density approximation (LDA) \cite{Dunjko}
(which requires sufficiently small $\omega_x$),
and, moreover, explicit results were only obtained in the TF and TG limits.
To check that the above picture is
consistent and robust,
 in this section we consider a trap potential of the form
$V(x,t)=-V_0\delta(x) \Theta(-t)$, 
assuming the circular
 atom waveguide in Fig.~\ref{fig1} to already contain
$N_0$ atoms in the absence of $V(x,t)$, so that $\rho_0=N_0/L$ 
is the density away from the trap at $t<0$.
 Let us then imagine that
$N\ll N_0$ additional atoms are injected ($N\propto V_0$ is tuned by the
trap depth)
in the distant past at $x=0$ by adiabatically switching on $V(x,t)$,
e.g.~from a BEC atom reservoir using a quantum tweezer  \cite{Tweezer}.  
At time $t=0$, the trap is switched off, and
we again wish to compute the interference signal.  This setup
may be of relevance to experiments using injection of particles
with a quantum tweezer, or for microchip-trapped
cold atoms  where confinement potentials are quite steep
\cite{Hansel}.  Importantly, under the condition $N\ll N_0$,
this interference problem can be solved exactly by virtue of the
bosonization method.  Using Eq.~(\ref{density-phase})
and the quantum fluctuations defined in Sec.~\ref{sec2}, one now arrives 
at a standard (uniform and time-independent) Luttinger liquid Hamiltonian
\cite{Monien} plus the contribution of the trap,
\begin{equation} \label{Hlas}
H(t)=  \frac{\hbar u}{2\pi } \int dx    [ K
 (\partial_x\Phi)^2 +  K^{-1} (\pi \Pi)^2 
]- V_0 \Theta(-t)  \Pi(0,t),
\end{equation}   
where $u$ is the sound velocity and 
$1\leq K <\infty$ is the dimensionless Luttinger parameter. Both
parameters are determined
by $g$ and $\rho_0$ \cite{Monien}.  In particular, in the TG limit, $K=1$,
while $K\gg 1$ in the TF limit.
The effect of the trap potential 
can then be included exactly by the time-dependent unitary
transformation  $\Pi(x,t)\to \Pi(x,t)+ N \delta(x)\Theta(-t)$,
where $N= KV_0/\pi\hbar u$ 
($N$ is assumed to be integer) gives the number of added atoms.
This allows us to obtain the expansion dynamics $\rho_0(x,t)=
\rho_0+\langle\Pi(x,t)\rangle$ and 
the interference signal in closed form. From the equation of motion,
\begin{equation}
(-\partial_t^2 + u^2 \partial_x^2) \Pi(x,t) =  N \Theta(-t)\partial_x^2 
\delta(x) ,
\end{equation}
for $t<0$ we infer the density $n(x) = \rho_0+  N \delta(x)$, while for
$t>0$ the expanding density profile is
\begin{equation}
\rho_0(x,t) = \rho_0+ N [ \delta(x-ut) + \delta(x+ut) ]/2.
\end{equation} 
Apparently, the initial $\delta$-peak in the density is
split into two counterpropagating parts, each moving with 
velocity $u$.  Furthermore, the phase $\phi_0(x,t)$ is easily found
in the form
\[
\phi_0(x,t)= \frac{\pi N}{2K} \Theta(t-|x|/u).
\]

The interference signal then follows from Eq.~(\ref{signal}),
 with the quantum fluctuations at $x\approx x'\approx \pm L/2$
 entering as 
\begin{equation}\label{FLL}
F(x,x',t) = \frac{1}{2K}
 \ln\left|\frac{\hbar u}{\pi a_0 k_B T}\sinh\left(\frac{ \pi L k_B T}
{\hbar u} \right)\right|,
\end{equation}
where $a_0$ is a nonuniversal
UV cutoff length limiting  the applicability of the continuum model
(\ref{Hlas}).  Note that due to our assumption of 
a constant density $\rho_0$ in the absence of the axial trap,
the fluctuation factor is now time-independent.
Equation (\ref{FLL}) implies power-law suppression of the
interference signal $\propto 1/L^{1/2K}$ for short circumference
$L<L_T$, where 
\begin{equation}\label{LT}
L_T=  \hbar u / \pi k_B T.
\end{equation}
This power law decays quite rapidly 
in the Tonks limit ($K=1$),  but interference
is only weakly suppressed in the TF limit.
Moreover, for large circumference $L>L_T$,
we find an exponential suppression of the interference signal, 
$\propto \exp(- L/2 K L_T)$.
The findings of the previous section are therefore confirmed from
this simple yet exact calculation.  There is a smooth crossover
between the TF and TG limits, described by the above expressions.

\section{Conclusions}\label{sec4}

In this paper, interference properties of interacting atomic Bose
gases in a 1D circular trap have been analyzed.  After switching 
off a trapping potential keeping the atoms in the narrow initial
region, the expanding atom clouds meet at the opposite side and
may produce an interference signal.   This signal has been 
computed for the case of a harmonic initial trap, with explicit
results in the Tonks-Girardeau regime and in the Thomas-Fermi limit,
including the effects of thermal fluctuations.
Our central conclusion is that many-body interactions strongly
suppress phase coherence and interference.  In fact, using this
setup interference is unlikely to be observed in the strongly
interacting Tonks gas.  Even in the Thomas-Fermi regime, 
one needs rather low temperatures and steep initial
traps to ensure a significant interference signal.
Further support for these conclusions can be drawn from
a simple yet exact calculation based on a slightly different
setup where one injects a small number of additional atoms
into the system, traps them for some time and then studies
the expansion and the subsequent interference signal.  

\acknowledgments

We thank R. Graham for valuable discussions. 
This work has been supported by the DFG under the SFB/TR-12.

\end{document}